\documentclass[aps,amssymb,amsmath,superscriptaddress,pra,twocolumn,showpacs]{revtex4-1}
\usepackage{graphicx}
\usepackage{subfigure}
\usepackage{amsfonts}
\begin{document}
\title{Entanglement entropy and fidelity susceptibility in the one-dimensional spin-1 XXZ chains with alternating single-site anisotropy}
\author{Jie Ren}
\affiliation{Department of Physics and Jiangsu Laboratory of Advanced
Functional Material, Changshu Institute of Technology, Changshu 215500, China}

\author{Guang-Hua Liu}
\affiliation{Department of Physics, Tianjin Polytechnic University, Tianjin 300387, People¡¯s Republic of China}

\author{Wen-Long You}
\affiliation{College of Physics, Optoelectronics and Energy, Soochow University, Suzhou, Jiangsu 215006, People¡¯s Republic of China}
\date{\today}

\begin{abstract}
 We study the fidelity susceptibility in an antiferromagnetic spin-1  XXZ chain numerically. By using the density-matrix renormalization group method, the effects of the alternating single-site anisotropy $D$ on fidelity susceptibility are investigated. Its relation with the quantum phase transition is analyzed. It is found that the quantum phase transition from the Haldane spin liquid to periodic N\'{e}el spin solid can be well characterized by the fidelity. Finite size scaling of fidelity susceptibility shows a power-law divergence at criticality, which indicates the quantum phase transition is of second order. The results are confirmed by the second derivative of the ground-state energy. We also study the relationship between the entanglement entropy, the Schmidt gap and quantum phase transitions. Conclusions drawn from these quantum information observables agree well with each other.
\end{abstract}
\pacs{03.67.-a,05.30.Jp}
\maketitle

\section{introduction}
\label{sec:intorduction}
In the last two decades, quantum phase transitions (QPTs) in quantum spin chains have attracted considerable interest both in experimental and theoretical research\cite{Sachdev}.
Among them, the spin $S$ = 1 antiferromagnetic Heisenberg chain has been extensively studied by many authors\cite{Nijs,Chen,Boschi}, whose ground state is termed the Haldane phase. Such phase has a peculiar nonlocal string order in which the spins with $\vert S_z=\pm 1\rangle$ are arranged antiferromagnetically if the sites with $\vert S_z= 0 \rangle$ are skipped. As we know, the QPT has been traditionally described based on the behavior of expectation values of local operators and two-point correlators in accordance to the ¡°standard¡± Ginzburg-Landau theory. Unlike the case of spontaneous symmetry breaking, the absence of the long-range order in Haldane phase is accompanied by a hidden $Z_2 \times Z_2$ symmetry breaking. Furthermore, the Haldane state is gapped between a spin-singlet ground state and a spin-triplet excited state, indicating spin-1 Heisenberg model is sharply different from spin-1/2 counterpart.

Experimentally, a few quasi-one-dimensional Haldane chain compounds have been investigated, such as
Y$_2$BaNiO$_5$\cite{Darriet},CsNiCl$_3$\cite{Buyers,Steiner,Morra}, Dy$_2$BaNiO$_5$\cite{Singh}, Nd$_2$BaNiO$_5$ \cite{Zheludev96,Raymond} were considered to realize $S$ = 1 Haldane systems with a magnetic gap in the excitation spectra. However, an ideal one-dimensional (1D) spin-1 system is rare in real materials, usually followed by the interchain interactions and magnetic anisotropy, which may partially or completely suppress the excitation gap and thus lead to an observation of long-range order in a quantum disordered magnet. The strength of single-ion anisotropy $D$ was retrieved from inelastic neutron-scattering on quasi-one-dimensional spin-1 chain compound ANi$_2$V$_2$O$_8$ (A =Pb and Sr)\cite{Zheludev}, electron spin resonance (ESR) study on  PbNi$_2$V$_2$O$_8$\cite{Smirnov}, multifrequency ESR transmission spectroscopy on single crystals SrNi$_2$V$_2$O$_8$\cite{Wang,Bera}. With large single-site anisotropy ($D$$>$0), the Haldane ground state changes to the large-$D$ state without explicit order, where  all spins are confined to the configurations $\vert S_z= 0 \rangle$.
On the other hand, the reverse single-site anisotropy $D$ ($D$$<$0) also drives the Haldane state into the N\'{e}el state. Note that in recent years oxide heterostructure appears to be an appealing discovery platform for emergent electronic states\cite{Chakhalian}. More explicitly, by suitable mechanical, electrical, or optical boundary conditions at oxide interfaces, intersite couplings and novel magnetic states can be manipulated. For instance, a possible scenario for realization of 1D zigzag chain in layered structures of transitional metal oxide was proposed\cite{You14}. The reduced symmetry at the interface strongly modifies the magnetocrystalline anisotropy at the interface. As a consequence, amplitudes and even the signs of the effective
single-ion anisotropy can fluctuate. Hence the large interest arises when considering the anisotropy effects, that is, the exchange anisotropy and the easy-plane staggered single-ion anisotropy coexist in spin-1 chains.

Recently, various exogenous approaches inherited from
quantum information to understanding quantum many-body systems exploit the curvature measures of the ground states. Much effort focused on the study of QPTs in spin chains in terms of entanglement entropy (EE)\cite{Amico}. Another concept referred to simply as fidelity susceptibility (FS),
which measures the changing rate between two states, will diverge at
the critical points\cite{Gu}. The ground-state EE and the FS were deemed plausible to qualify QPTs in strong correlated systems \cite{Abasto,Buonsante,You,You2,Zhou01,legeza,Ren01,Ren02,You12,Liu12,Ren03}. It is due to QPTs are intuitively accompanied by an abrupt change in the structure of the ground-state wave function. This primary observation motivates researchers to use the EE and the FS to predict QPTs. The scaling relation of the FS and the EE proposed recently is verified for the spin-1 XXZ spin chain with a
 single-site anisotropy term\cite{Tzeng}. Through a proper finite-size scaling analysis, the results from both the FS and the EE, agree with the findings in the previous results\cite{Tzeng1}. The effect of spatial modulation of site-dependent anisotropy $D_i$ in the $S$ = 1 Heisenberg chain was studied using perturbation theory and exact diagonalization on small-size system\cite{Hida}. Thus, in order to figure out the QPT in 1D spin-1 Heisenberg chains with exchange anisotropy and alternating single-site anisotropy, it is better to consider these two quantum information observables again.

In the present paper, we calculate the ground-state FS, the EE and the Schmidt gap in 1D spin-1 XXZ chains with alternating single-site anisotropy, and make them ideal tools for searching the phase transition points. The paper is organized as following. In Sec \ref{sec:intorduction}, an introduction is provided. The Hamiltonian is shown in Sec \ref{sec:Hamiltonian}. The measurements and details of method to obtain the ground-state properties are introduced in Sec. \ref{sec:Algorithm}. In Sec \ref{sec:results}, results of the fidelity calculation and the EE as well as the Schmidt gap of the system are presented. A discussion is provided in the last section.
\section{Hamiltonian}
\label{sec:Hamiltonian}
The Hamiltonian of a 1D spin-1 XXZ chain with alternating single-site anisotropy is given by
\begin{eqnarray}
\label{eq1}
&H=
\displaystyle{  \sum_{i=1}^{N}}J(S_i^x S_{i+1}^x+S_i^y S_{i+1}^y+\lambda S_i^z S_{i+1}^z)+ (-1)^iD(S_i^z)^2, \nonumber \\
\end{eqnarray}
where $S_i^\alpha(\alpha=x,y,z)$ are spin-$1$ operators on the $i$-th site and $N$
is the length of the spin chain. The periodic boundary condition is considered, and it is denoted that $N+1=1$.
The parameter $J$ denotes the antiferromagnetic coupling, and $J$=1 is considered in the paper. The parameters $\lambda$ and $D$ are the anisotropic spin-spin interaction and single-site anisotropy, respectively.

\section{Measurements and Algorithm }
\label{sec:Algorithm}
A concept from quantum information theory, i.e., the ground-state
fidelity, can be applied to capture the occurrence of the
QPTs. Taking a general Hamiltonian $
H(D)=H_0+D H_I $ as an example, where $H_0$ is the main
part, $H_I$ is the driving part and the quantum parameter
$D$ denotes its strength. If $\rho(D)$ represents a density matrix
of the system, the ground-state fidelity between
$\rho_0(D)$ and $\rho_0(D+\delta D)$ can be
defined as
\begin{equation}
\label{eq2}
F(D,\delta D)=Tr[\sqrt{\rho_0^{1/2}(D)\rho_0(D+\delta D)\rho_0^{1/2}(D)}],
\end{equation}
where $\delta D$ is a small deviation. For a pure state
$\rho_0=|\psi_0\rangle\langle \psi_0|$, Eq. (\ref{eq2}) can be
rewritten as
\begin{equation}
 F(D,\delta D)=|\langle
\psi_0(D)|\psi_0(D+\delta D)\rangle|,
\end{equation}
which
represents the overlap of the wavefunctions at two adjacent
quantum parameter points, and $F(D,\delta D)$ reaches
its maximum value $F_{max}=1$ at $\delta D=0$. Expanding
$|\psi_0(D+\delta D)\rangle$ to the first order, we have
\begin{equation}
\label{eq3}|\psi_0(D+\delta D)\rangle=|\psi_0(D)\rangle+\delta D \sum_{n\neq0}\frac{H_{n0}|\psi_n
(D)\rangle}{E_0(D)-E_n(D)},
\end{equation}
where $H_{n0}=\langle \psi_n(D)|H_I|\psi_0(D)\rangle$,
and the eigenstates $|\psi_n(D)\rangle$ satisfy
$H(D)|\psi_n(D)\rangle=E_n|\psi_n(D)\rangle$.
Therefore, the fidelity susceptibility removes the artificial variable $\delta D$~\cite{Buonsante,You} and can be calculated by
\begin{equation}
\label{eq5}
\chi_F(D)=\lim_{\delta D\rightarrow
0}\frac{-2 \textrm{ln}F(D,\delta D)}{(\delta D)^2},
\end{equation}
and then it yields
\begin{equation}
\chi_F(D)=\displaystyle{\sum_{n\neq0}}\frac{|\langle\psi_0(D)|H_{I}|\psi_n(D)\rangle|^2}{[E_0(D)-E_n(D)]^2}.
\end{equation}
The divergence of FS can directly locate the critical points. The related phase transition is exactly convincing\cite{Cozzini,You,You2}.

The EE can be chosen as a measurement of the bipartite
entanglement, which is defined as follows. Assuming $|g.s.\rangle$ is
the ground state of the target Hamiltonian, which can be divided into two subsystems $A$ and $B$.
One convenient choice of subsystem $A$ is composed from the first site to the $L$th site and the subsystem $B$
is the rest of the system. The reduced density matrix of
part $L$ can be obtained by taking the partial trace over system
$N-L$, which is given by
\begin{equation}
\label{eq6}\rho_{A}=Tr_{B}(|g.s.\rangle \langle g.s.|).
\end{equation}
Then, the bipartite EE measures the entanglement between parts $A$ and $B$ as

\begin{equation}
\label{eq7}S_L=-Tr(\rho_{A}\log_2\rho_{A}).
\end{equation}
In addition, as a local order parameter, the Schmidt gap can also be used to describe the QPTs\cite{Chiara}. It is defined as
\begin{equation}
\label{eq8}G=g_1-g_2,
\end{equation}
where $g_1$ and $g_2$ are the first and the second largest eigenvalues of the reduced density matrix $\rho_{A}$ [Eq. (\ref{eq6})], respectively.

Thanks to the density-matrix
renormalization-group(DMRG)~\cite{white,U01,U02} method, the
ground state of the 1D system can be calculated with
very high accuracy. We implement GPU speeding up Matlab
code for the finite-size DMRG with double precision.
The maximum eigenstates kept is
$m=200$ during the procedure of basis truncation, and the truncation error is smaller than $10^{-8}$ for system sizes up to
$N=100$. With such high performance calculation, we can precisely
analyze the QPTs in terms of both the EE and the FS.

\section{Numerical results}
\label{sec:results}

\begin{figure}[t]
\includegraphics[width=0.40\textwidth]{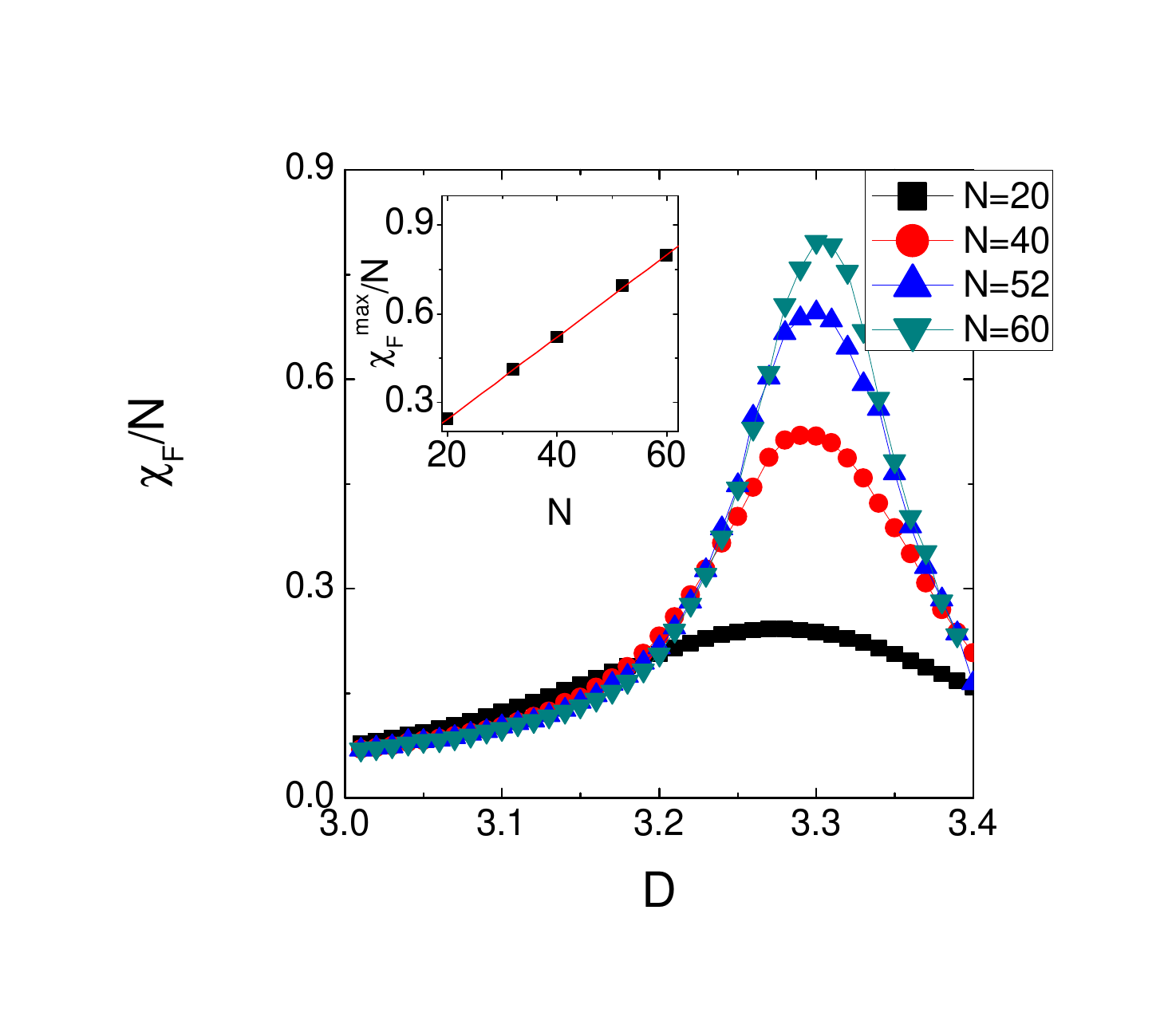}
\caption{(Color online) The fidelity
susceptibility per site is plotted as a function of the single-site
interaction for different system sizes with $\lambda=1$. Inset: $\chi_F^{max}/N$ for various sizes $N$. The line is a linear fit for guiding eyes.}
\label{fig6}
\end{figure}

As a check, we plot the FS per site as a function of the single-site
interaction for different system sizes with $\lambda=1$. As shown in the Fig. \ref{fig6}, peaks in the ground-state FS are observed, which signal precursors of the phase transition from Haldane phase to periodic N\'{e}el spin solid at $D_c=3.30$, and this was confirmed by previous results\cite{Hida}.

\begin{figure}[t]
\includegraphics[width=0.40\textwidth]{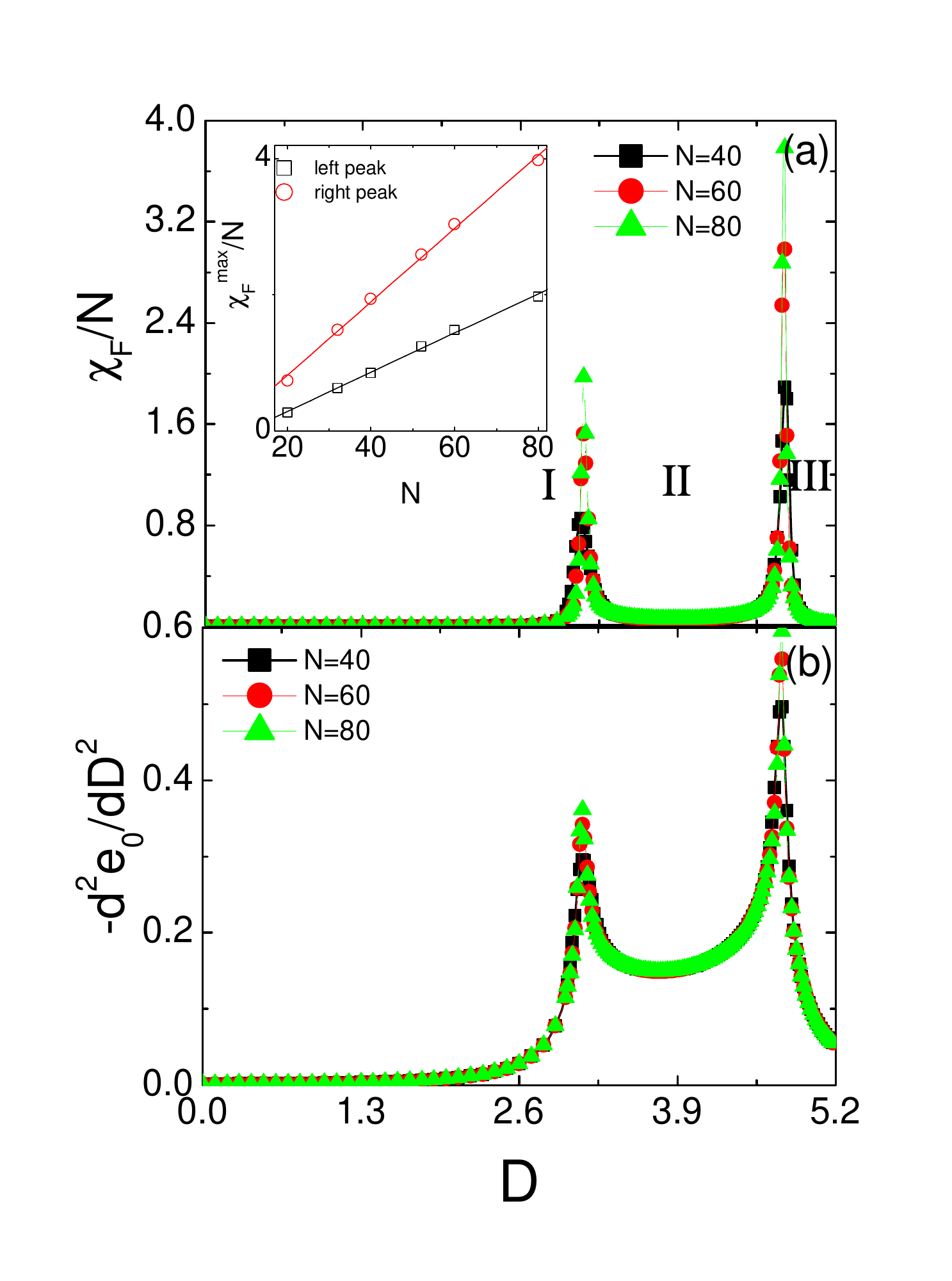}
\caption{\label{fig1} (Color online) (a) The fidelity
susceptibility per site is plotted as a function of the single-site
anisotropy $D$ for different system sizes with $\lambda=2$. Inset: $\chi_F^{max}/N$ for various sizes $N$. The lines are fitting lines. (b) The second derivative of the ground-state energy density is plotted as a function of the single-site interaction for different system sizes with $\lambda=2$.}
\end{figure}
\begin{figure}
\includegraphics[width=0.40\textwidth]{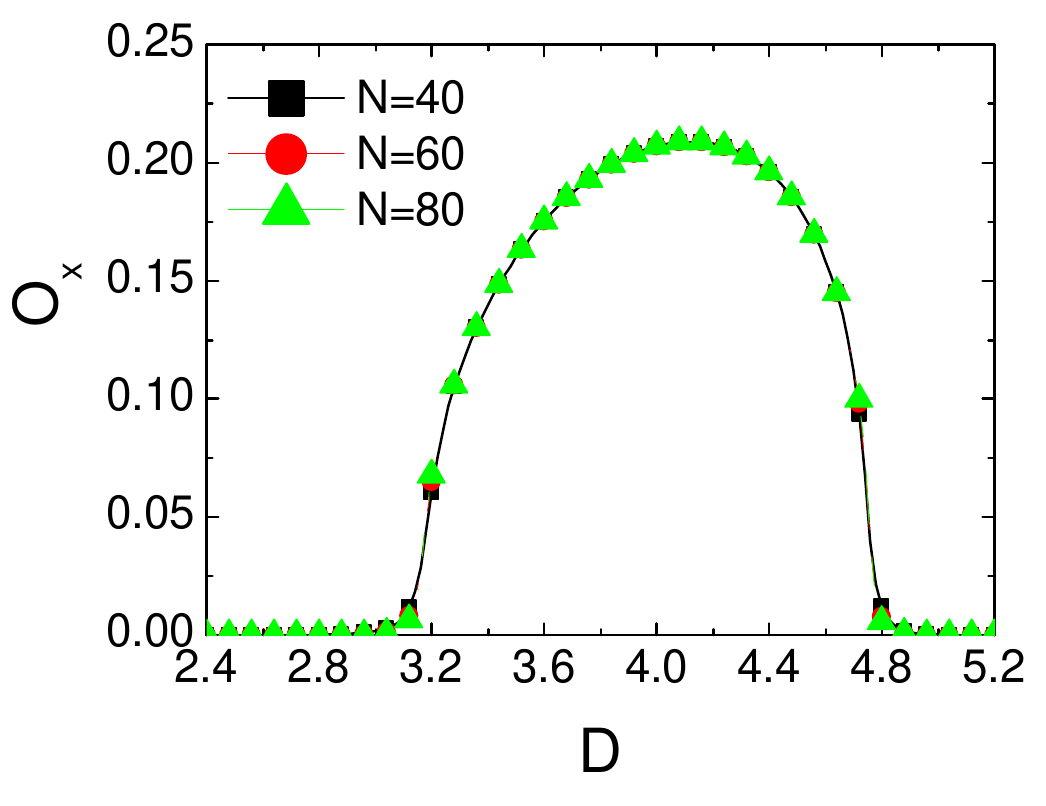}
\caption{\label{fig2} (Color online)The string order parameter $O_x$ is plotted as a function of the single-site anisotropy for different system sizes with $\lambda=2$.   }
\end{figure}

In Fig. \ref{fig1}(a), we plot the FS per site of XXZ chain as a function
of single-site anisotropy $D$ for different lattice sizes $N$=40, 60, 80.
Two peaks of the ground-state FS are observed, indicating that the dramatic changes in the ground-state structure with the increase of $D$ take place twice. Initially, i.e., $D=0$, the system is in N\'{e}el state
$\vert \uparrow \downarrow \uparrow \downarrow \cdots \uparrow \downarrow \rangle$ when $\lambda=2$. The system becomes periodic  N\'{e}el state $\vert \uparrow 0 \downarrow 0 \uparrow 0 \downarrow \cdots 0 \uparrow 0\downarrow \rangle$ when $D$ is large enough. In order to characterize the intermediate phase,
we calculate the string order parameter (SOP), whose definition
is given by\cite{Nijs}
\begin{equation}
O_{x} = - \lim_{j-i \to \infty} [S_i^{x} \exp(i \pi \sum_{i<l<j} S_l^{x}) S_j^{x}].
\label{eq9}
\end{equation}                                                                                                      The SOP characterizes the topological order in the Haldane phase\cite{Ueda,Wei}. The results are shown in Fig.\ref{fig2}, where $O_{x}$ is nonzero in the intermediate phase, implying it is the Haldane phase.
The finite-size scaling of the FS shows a power-law divergence at criticality, and this suggests that both QPTs belong to a second-order transition\cite{You2}. In Fig. \ref{fig1}(b), we also study the second derivative of the ground-state energy density, which exhibits divergence  at criticalities. This confirms that the transitions should be of second order.
\begin{figure}[t]
\includegraphics[width=0.450\textwidth]{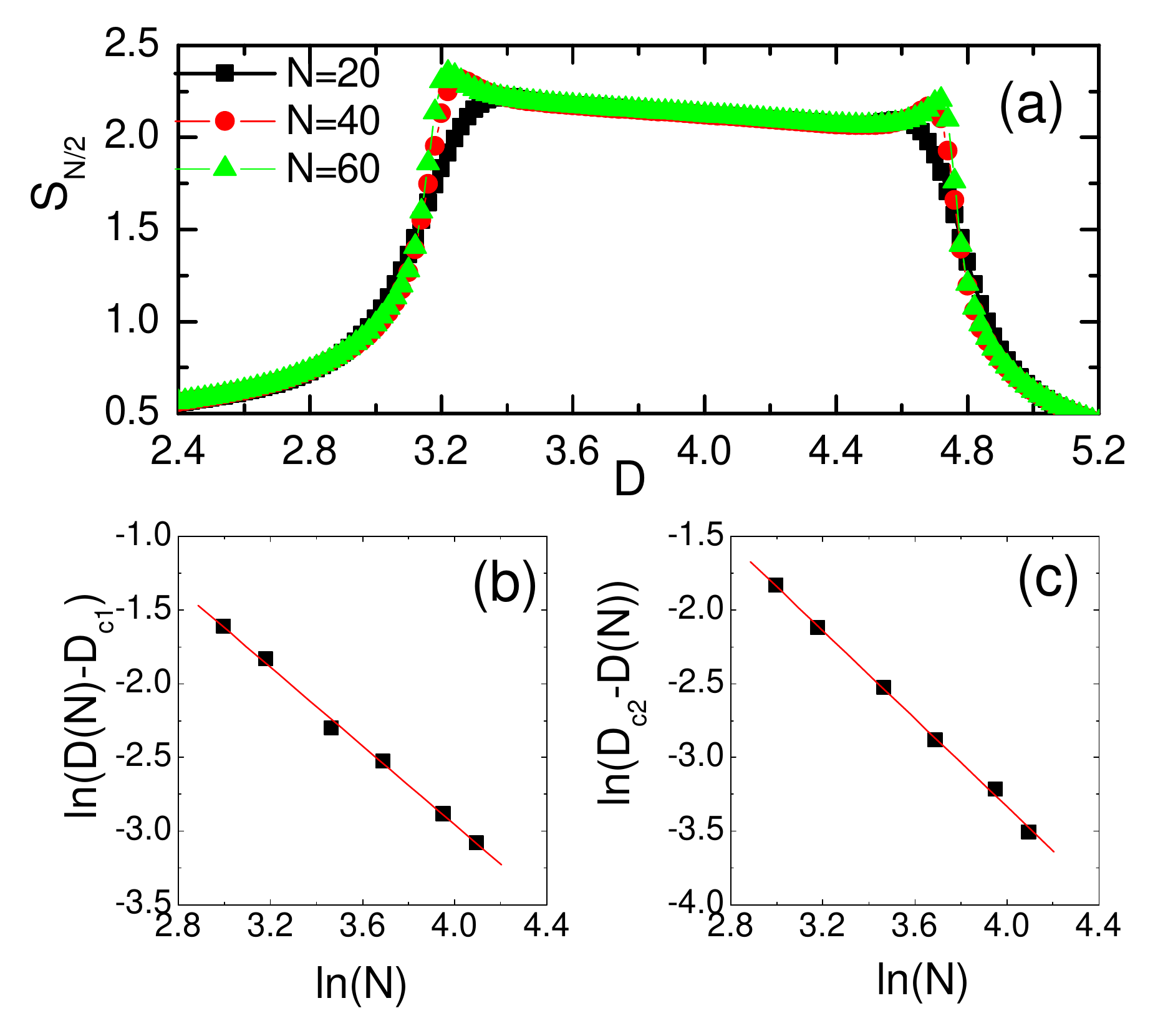}
\caption{\label{fig3} (Color online)(a)The block entanglement between two half subsystems labeled by $E$ is plotted as a function of D for different system sizes with $\lambda=2$.(b),(c) Finite-size scaling of $D_{c}$ of the EE. The lines are the fit lines.}
\end{figure}

In order to locate the QPTs precisely, the EE
between two half subsystems is plotted as a function of $D$ for different system sizes in Fig. \ref{fig3}(a). It is observed that starting from small $D$ the entanglement firstly grows as the $D$ increases, meaning the quantum fluctuation enhance the quantum nature in the classical-like system. After reaching a local maximum, the variation of the EE becomes relatively small. When the $D$ increases further, the EE reaches another maximum and then decreases rapidly. Furthermore, the EE is insensitive to the system size, as a consequence of boundary law in 1D gapped Hamiltonian, while the maximum values of the EE grow with increasing $N$, which is a logarithmically divergent correction to the boundary law\cite{Wolf2,Gioev}. The location of left peak in Fig.\ref{fig3}(a) moves to higher $D$ up to a particular value as the system size increases, while the location of right peak moves to lower $D$ up to a particular value as the system size grows. We fit the locations of maximums by the formula
\begin{eqnarray}
D_c(N)\sim D_c+aN^{-b},
\end{eqnarray}
where $a$, $b$ are size-independent constants and $N$ is the system size. We obtain that $D_{c1}=3.20$, $a_1=14.44$, $b_1=1.438$ and $D_{c2}=4.760$, $a_2=-14.45$, $b_2=1.506$; see Fig. \ref{fig3}(b) and (c).

\begin{figure}[t]
\includegraphics[width=0.500\textwidth]{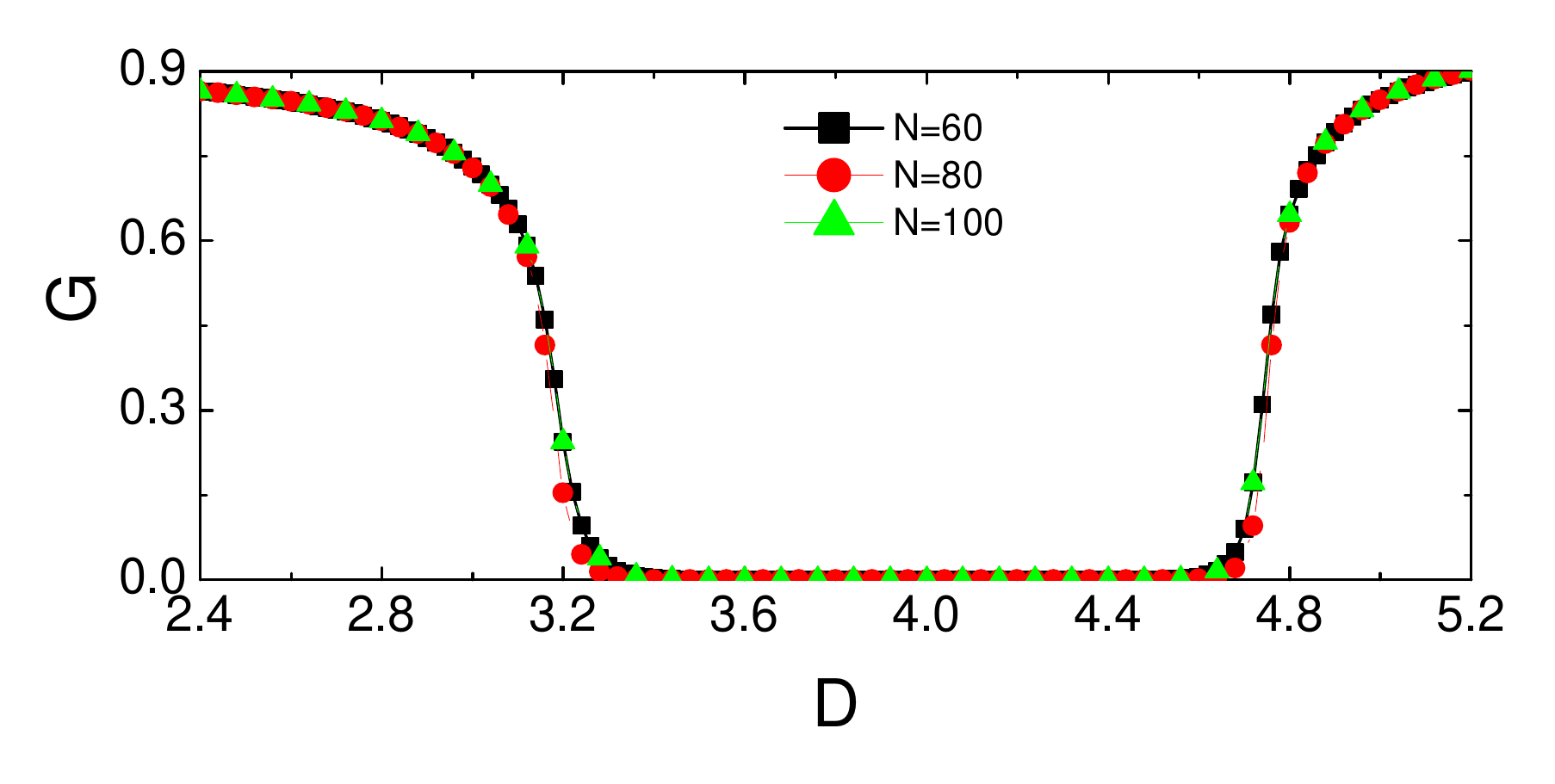}
\caption{\label{fig4} (Color online)The Schmidt gap labeled by $G$ is plotted as a function of D for different system sizes with $\lambda=2$.}
\end{figure}

\begin{figure}[t]
\includegraphics[width=0.35\textwidth]{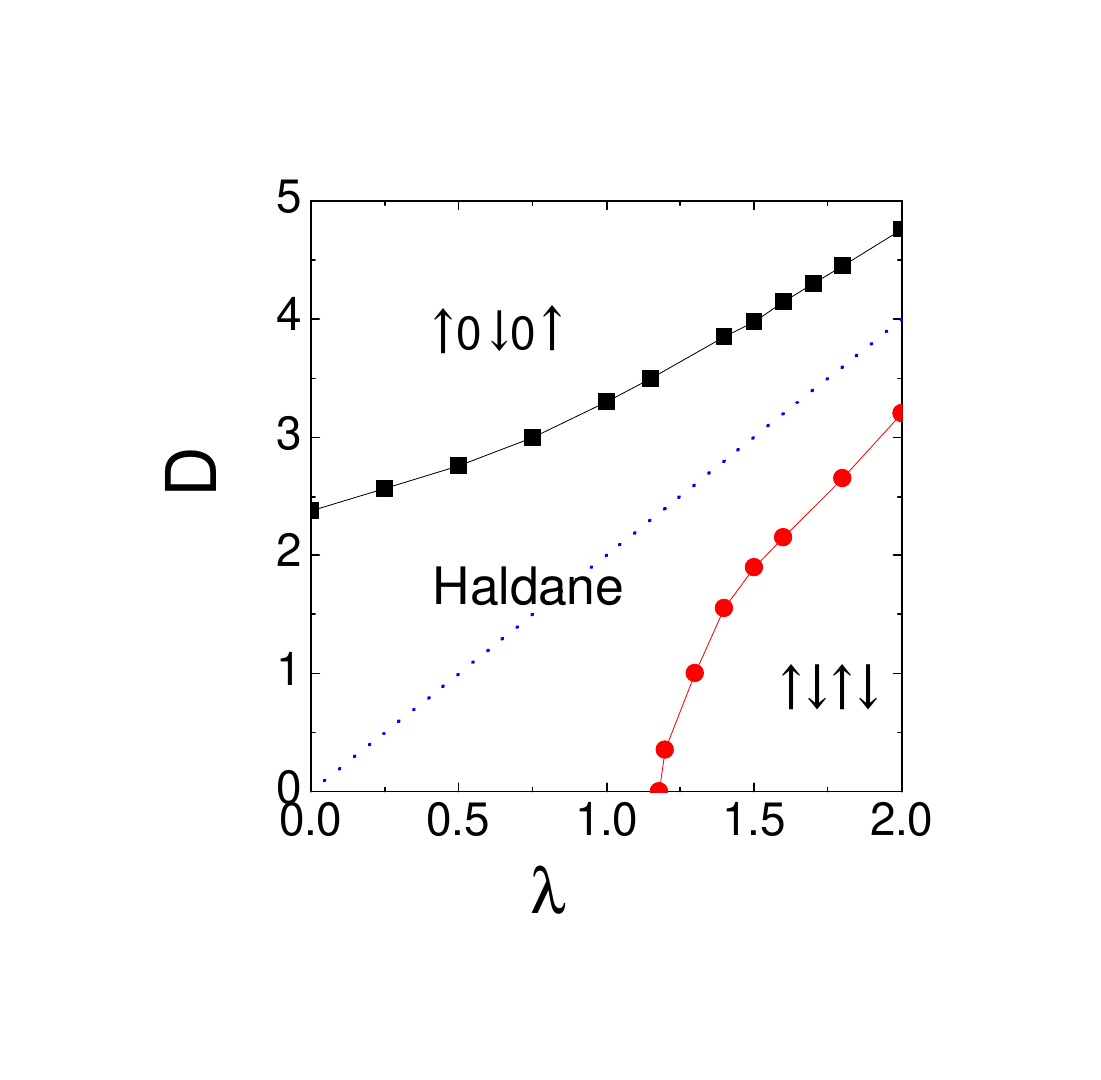}
\caption{\label{fig5} (Color online)Phase diagram of a spin-1 Heisenberg chain with alternating anisotropy $D$ and $\lambda$. Here $\uparrow$, $0$ and $\downarrow$ stand for the single-site states $|S_i^z\rangle$ with $S_i^z = 1$, 0 and -1 respectively. $\uparrow \downarrow \uparrow \downarrow \uparrow \downarrow $ ($\uparrow 0 \downarrow 0\uparrow 0 \downarrow$ ) denotes the N\'{e}el (periodic N\'{e}el) phase. The dot line $D=2\lambda$.}
\end{figure}

We also investigate the Schmidt gap of the reduced density matrix after cutting a $N$-site chain into two half subsystems $(L=N/2)$. The Schmidt gap labeled by $G$ is plotted as a function of $D$ for different system sizes in Fig. \ref{fig4}. It is seen that the Schmidt gap is large when the system is in the N\'{e}el phase. With increasing $D$, the Schmidt gap closes very rapidly when the system is in the Haldane phase, because S=1 Haldane phase is a topological
phase protected by specific global symmetry and is characterized by a double degeneracy of the entanglement
spectrum\cite{Pollmann}. When $D$ increases further, the Schmidt gap opens very rapidly again when system is in the periodic N\'{e}el phase.

In Fig. \ref{fig5}, we portray the $D$-$\lambda$ phase diagram of the Hamiltonian (\ref{eq1}), which is detected via the EE, the Schmidt gap and the FS. When $D=0$, the quantum transition from Haldane spin liquid to N\'{e}el spin solid is continuous and belongs to a second-order QPT at $\lambda=1.18$\cite{Chen,Ren01}. The critical points between the periodic N\'{e}el phase and the Haldane phase decrease with $\lambda$ increases. The critical points between the N\'{e}el phase and the Haldane phase increase as $\lambda$ increases. Two critical lines will merge each other when $\lambda$ increases, where the Haldane phase disappears and the N\'{e}el$-$periodic N\'{e}el transition becomes first order at $D=2\lambda$.

\section{Discussion}
\label{sec:Discussion}
In the present paper, we have investigated the QPTs in the 1D spin-1 XXZ chains with alternating single-site anisotropy by analysing the bipartite entanglement, the Schmidt gap and the FS by using the DMRG technique. Their relation with QPTs is under discussion. It is important to note that the quantum phase transitions from the N\'{e}el ordering to Haldane spin liquid to periodic N\'{e}el spin solid can be well characterized by the FS. The finite-size scaling demonstrates that FS should diverge in the thermodynamic limit at the pseudo-critical point, and the locations of extreme points approach quantum critical point accordingly. It also shows a power-law divergence at criticality, which indicates the QPT is of second order, and the critical exponent can be obtained. We compare the FS with the second derivative of the ground-state energy, and find both of them exhibit similar peaks. The critical point can also be successfully detected by the EE and the Schmidt gap. To sum up, the quantum information observables are effective tools for detecting diverse QPTs in spin-1 models.

\vspace{0.3cm}
\begin{acknowledgments}

This work is supported by the National Natural Science
Foundation of China (NSFC) under Grants No. 11104021,  No. 11347008 and No. 11474211, as well as the Natural Science Foundation
of Jiangsu Province of China under Grant No. BK20141190.
\end{acknowledgments}

\end{document}